\documentclass{iau}
\usepackage{graphicx}
\usepackage{natbib}

\newcommand{\apj}{\textit{ApJ}}
\newcommand{\apjl}{\textit{ApJ.\ Lett.}}
\newcommand{\apjs}{\textit{ApJS}}
\newcommand{\prd}{\textit{Phys.\ Rev.\ D}}
\newcommand{\prl}{\textit{Phys.\ Rev.\ Lett.}}
\newcommand{\nat}{\textit{Nature}}
\newcommand{\jcap}{\textit{JCAP}}
\newcommand{\physrep}{\textit{Phys.\ Rep.}}
\newcommand{\vnhat}{{\hat{\mathbf{n}}}}
\newcommand{\vb}{\mathbf{v}_b}

\title[CMB anisotropy science: a review] 
{CMB anisotropy science: a review}

\author[Anthony Challinor]
{Anthony Challinor} 

\affiliation{
Institute of Astronomy and Kavli Institute for Cosmology Cambridge,
Madingley Road, Cambridge, CB3 0HA, U.K. \\[\affilskip]
DAMTP, Centre for Mathematical Sciences, Wilberforce Road, Cambridge,
CB3 0WA, U.K. \\
email:{\tt a.d.challinor@ast.cam.ac.uk}}

\pubyear{2012}
\volume{288}  
\pagerange{119--126}
\jname{Astrophysics from Antarctica}
\editors{M.G. Burton, X. Cui \& N.F.H. Tothill, eds.}
\begin{document}

\maketitle

\begin{abstract}
The cosmic microwave background (CMB) provides us with our most direct observational window to the early universe.
Observations of the temperature and polarization anisotropies in the CMB have played a critical role in defining the now-standard cosmological model.
In this contribution we review some of the basics of CMB science, highlighting the role of observations made with ground-based and balloon-borne Antarctic telescopes. Most of the ingredients of the standard cosmological model are poorly understood in terms of fundamental physics.
We discuss how current and future CMB observations can address some of these issues, focusing on two directly relevant for Antarctic programmes:
searching for gravitational waves from inflation via $B$-mode polarization, and
mapping dark matter through CMB lensing.
\end{abstract}

\firstsection 
\section{Introduction}

It is now twenty years since the landmark discovery of fluctuations in the
temperature of the cosmic microwave background radiation by the COBE
satellite~\citep{1992ApJ...396L...1S}. Over the
intervening period, a now-standard cosmological model has emerged. The CMB fluctuations have been pivotal in putting this model on a firm
observational footing (though many of its key ingredients continue to defy
explanation in fundamental physics), and in measuring its parameters to a level of precision that is unprecedented in cosmology. While the game-changer in this field
has undoubtedly been the full-sky measurements from the WMAP satellite, observations of the CMB from Antarctica have played an important role in this development and
have achieved a number of significant `firsts'. These include
precision measurements of spatial flatness and the detection
and, now, characterisation of linear polarization of the CMB.

This symposium covers a broad range of astrophysics so the purpose of this
review is to set the scence for the other more specialised CMB contributions that follow.
We begin by reviewing some of the basics of CMB science and the
remarkable achievements made through observations of the CMB temperature and polarization fluctuations, highlighting the role of measurements made from Antarctica.
Experiments in Antarctica are also very much at the cutting edge of
future programmes seeking to address some of the questions raised by
the standard cosmological model. Space limits us to discuss in detail only
two of the main science goals of these experiments: the quest for gravitational
waves and CMB lensing. For more complete recent reviews of CMB science,
see~\cite{2009AIPC.1132...86C} and~\cite{2008arXiv0802.3688H}. 

\section{The CMB and the standard cosmological model}

In the standard cosmological model, named $\Lambda$CDM, the universe is well described
on large scales by a spatially-flat, homogeneous and isotropic background
metric with small fluctuations at the $10^{-5}$ level. The universe
has evolved from a hot, dense phase during which matter and radiation
were in thermal equilibrium at sufficiently early time. The CMB is the thermal
relic radiation from this early phase and its existence is a cornerstone
of the hot big bang model. The CMB radiation has now cooled to a temperature
of $2.725\,\mathrm{K}$ but retains an almost perfect blackbody spectrum.
Baryons and leptons make up $4.5\%$ of the current energy density and
cold dark matter (CDM; hypothesised matter with essentially only
gravitational interactions and negligible thermal velocities) $22\%$.
The remaining $73\%$ is in the form of dark energy and drives the current
accelerated expansion. Dark energy is not understood at all at a physical level but
phenomenologically behaves like a smoothly distributed fluid with 
equation of state close to $p = -\rho$, as for a cosmological constant $\Lambda$.

The flatness and large-scale smoothness of the universe are neatly explained
by a hypothesised period of quasi-exponential expansion -- cosmic inflation --
in the early universe. During a period of only $10^{-32}\,\mathrm{s}$,
the universe expanded in size by at least 60 e-folds. Inflation is
not understood at a fundamental level, but it can be realised in simple
models by a scalar field $\phi$ evolving slowly over a flat part of its
self-interaction potential $V(\phi)$. A compelling feature of inflation
is that it naturally provides a causal mechanism for generating
primordial curvature perturbations and gravitational waves with nearly
scale-free power spectra. Small-scale quantum fluctuations in light scalar
fields, and the spacetime metric, are stretched beyond
the Hubble radius during inflation to appear later as classical,
long-wavelength cosmological perturbations that seed the growth of large-scale
structure. In simple models (e.g.\
$V(\phi) \propto \phi^2$) inflation at energies $E_{\mathrm{inf}} \sim
10^{16} \,\mathrm{GeV}$ reproduces the observed level of perturbations.

\subsection{Temperature anisotropies}

The CMB carries an imprint of the primordial perturbations via small
temperature anisotropies at the $O(10^{-5})$ level.
The universe became transparent to CMB photons
around the time of recombination, when atomic hydrogen (and helium) first
formed. This defines a last-scattering surface centred on our current
location, and spatial fluctuations in the CMB energy density, bulk velocity and
gravitational potential over this surface project to give temperature
anisotropies in the CMB. In more detail, for curvature perturbations,
the fractional anisotropy $\Theta(\vnhat)$ along a direction $\vnhat$
at time $t_0$ is given approximately by
\begin{equation}
\Theta(\vnhat) = \Theta_0 + \psi - \vnhat \cdot \vb + \int_{t_*}^{t_0}
(\dot{\psi} +\dot{\phi}) \, dt \, .
\label{eq:deltat}
\end{equation}
Here, $\Theta_0$ is the fractional fluctuation in the CMB temperature on the
last-scattering surface (time $t_*$), $\vb$ is the baryon peculiar
velocity, and $\psi$
and $\phi$ are the gravitational potentials ($\phi=\psi$ in general
relativity when non-relativistic matter is dominant). Each term has a
simple physical interpretation: we see the intrinsic temperature fluctuation
$\Theta_0$, modified by the gravitational-redshifting effect of the
potential $\psi$ and the Doppler shift from scattering off moving matter.
The final \emph{integrated Sachs-Wolfe} term in Eq.~(\ref{eq:deltat})
involves the integral of the time derivatives of $\psi$ and $\phi$; if a
potential well is getting shallower in time (as happens during dark-energy
domination), photons receive a net blueshift in crossing the well and the
CMB appears hotter.
In practice, around 10\% of photons were re-scattered after the
universe reionized which, on all but the largest scales, reduces the \emph{primary
anisotropies} sourced around recombination by a factor $e^{-\tau}$, where
$\tau \approx 0.1$ is the Thomson optical depth.

The small amplitude of the temperature anisotropies means they can
be calculated very accurately with linear perturbation theory.
The fluctuations on the last-scattering surface are therefore a linearly-processed
version of the nearly scale-free
primordial perturbation. On scales large compared to the
Hubble radius at last-scattering, only gravity is important but
on smaller scales the acoustic physics of the primordial plasma and
photon diffusion dominate. Gravity-driven infall will tend to enhance
a positive density perturbation, but this is resisted by photon pressure setting
up acoustic oscillations in the plasma. The sine and cosine-like modes of oscillation
extrapolate back to decaying and constant modes at early times. Inflation is
democratic, putting equal power into each mode at generation, but
any decaying mode is totally negligible by the time the
acoustic oscillations begin. This process leaves only cosine-like oscillations
in the plasma, so that oscillations on all scales start off in phase. However,
different scales oscillate at different frequencies and scales which
have reached extrema of their oscillations by last-scattering have enhanced
power in the anisotropies on the corresponding angular scales. In this way,
the sound horizon $r_s(t_*)$, i.e.\ the (comoving) distance a sound wave can have
propagated by time $t_*$, introduces a preferred length scale to the
fluctuations. It is a fortunate coincidence that the corresponding angular scale
$r_s / d_A$ (where $d_A$ is the angular-diameter distance back to last-scattering)
is around $1^\circ$ and so straightforward to observe at frequencies around
$100\,\mathrm{GHz}$ where the CMB is brightest. 

\begin{figure}
\begin{center}
\includegraphics[width=3.4in,angle=-90]{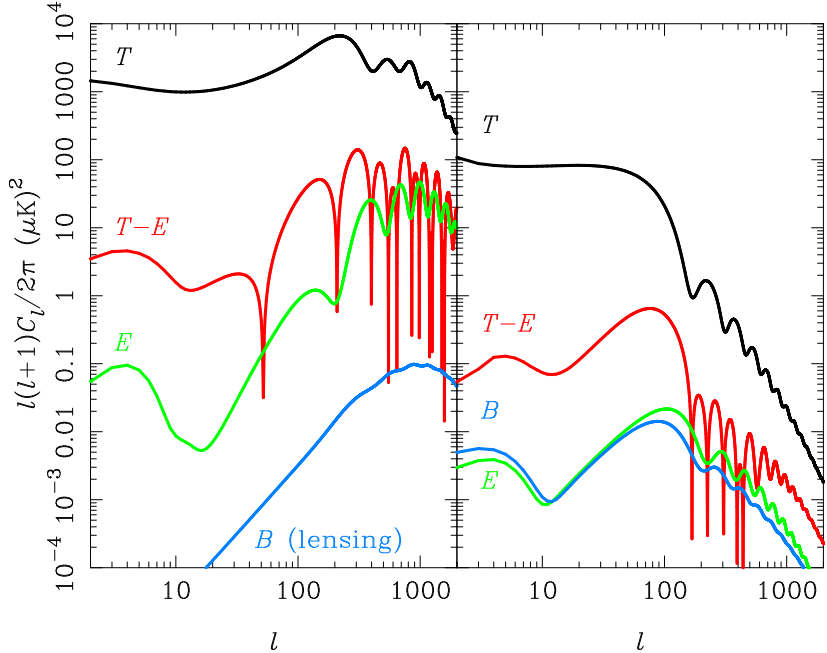} 
\caption{Temperature
(black), $E$-mode (green), $B$-mode (blue) and $TE$ cross-correlation
(red) CMB power spectra from curvature perturbations (left) and
gravitational waves (right) for a tensor-to-scalar ratio $r=0.24$.
The $B$-mode spectrum induced
by weak gravitational lensing is also shown in the left-hand panel (blue).}
\label{fig:scalar_tensor_spectra}
\end{center}
\end{figure}

Figure~\ref{fig:scalar_tensor_spectra}
shows the predicted angular power spectrum, $C_l^T$, from inflationary curvature perturbations. The power spectrum is the variance of the multipoles $\Theta_{lm}$ in a
spherical-harmonic expansion $\Theta(\vnhat) = \sum_{lm} \Theta_{lm} Y_{lm}(\vnhat)$.
The multipole index $l$ corresponds roughly to anisotropies at scale $180^\circ / l$.
The plateau in $C_l$ on large scales is from the combination of primary anisotropies
on scales large enough to be unaffected by acoustic processing, and from the integrated-Sachs-Wolfe effect from late-time decay of the gravitational potentials.
On intermediate scales, we have acoustic peaks. Finally, on smaller scales
the power decays rapidly. This \emph{damping tail} is sourced by perturbations
on scales small enough that photons had time to diffuse out of overdensities
by last-scattering, thus damping out the acoustic oscillations. This process imprints
another scale, the diffusion scale, into the CMB.

\begin{figure}
\begin{center}
\includegraphics[width=3.4in,angle=0]{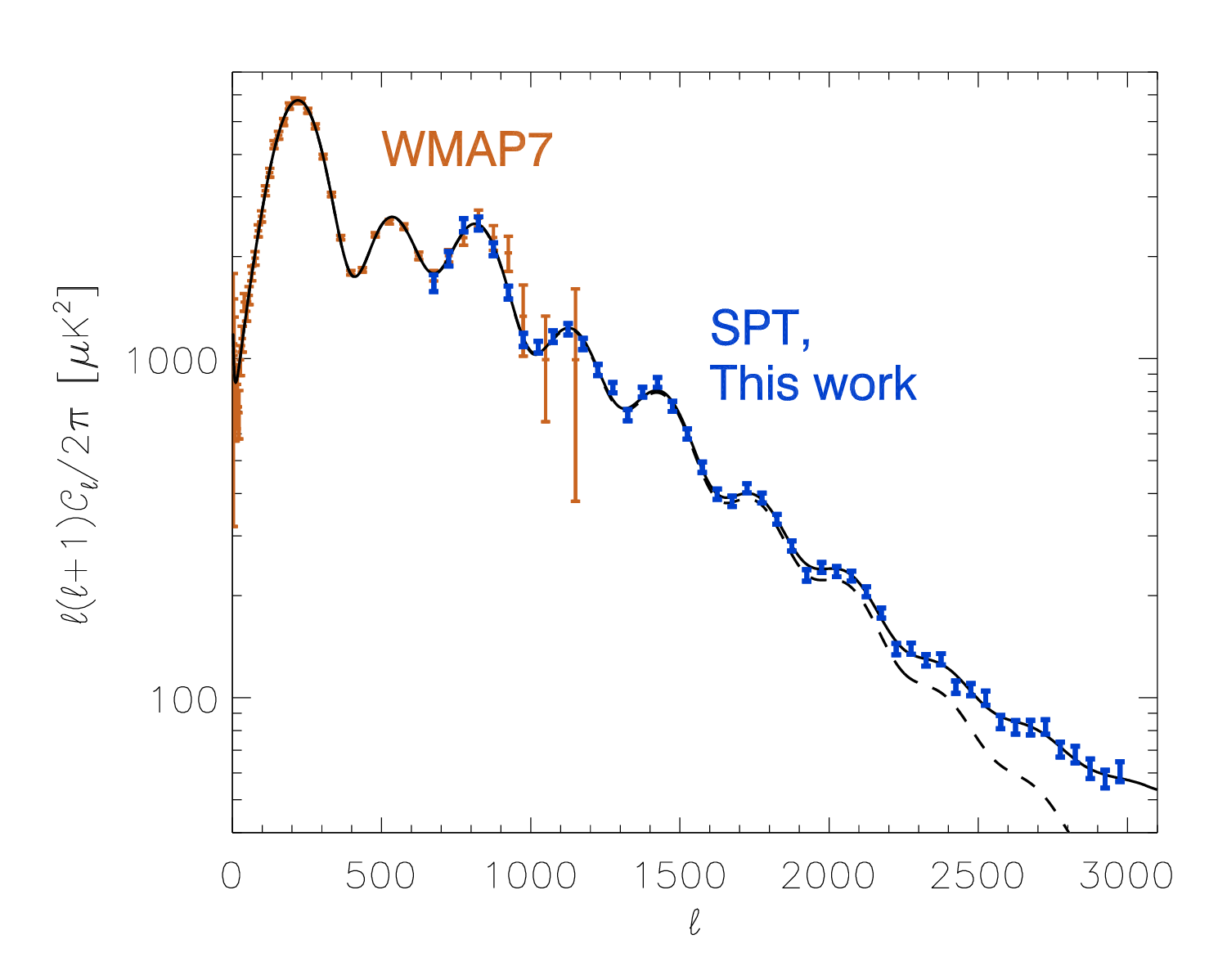} 
\caption{Measurements of the temperature power spectrum from WMAP (orange;
\citealt{2011ApJS..192...16L}) and $790\,\mathrm{deg}^2$ of the SPT 150-GHz survey (blue; \citealt{2011ApJ...743...28K}). Also shown are the CMB spectrum (dashed) and the
total spectrum (CMB and extragalactic foregrounds; solid) in the
best-fitting $\Lambda$CDM model. Reproduced with permission from~\cite{2011ApJ...743...28K}.
}
\label{fig:temperature_measurements}
\end{center}
\end{figure}

The picture outlined above is spectacularly confirmed by measurements of the
temperature anisotropy. Theory only allows us to predict the statistical properties
of the primordial perturbation. In simple models of inflation, the statistics are Gaussian,
and so fully characterised by their power spectrum. Primordial Gaussianity is borne out
by careful measurements of the statistics of the CMB anisotropies.
For this reason, the main
focus of observational CMB research for the past 20 years has been to obtain precise estimates
of the CMB angular power spectrum and to confront these against theoretical models.
An example of such measurements from the WMAP satellite~\citep{2011ApJS..192...16L} and
the South Pole Telescope (SPT;~\citealt{2011ApJ...743...28K}) is shown in Fig.~\ref{fig:temperature_measurements}. Nine acoustic peaks have now been
measured and the SPT measurements thoroughly characterise the damping tail.
The error bars on the power spectrum include the effects of instrument noise and
cosmic/sample variance -- at each $l$ we estimate the power spectrum from the empirical variance from a sample of only $(2l+1) f_{\mathrm{sky}}$ independent quantities
(where $f_{\mathrm{sky}}$ is the sky fraction covered by the survey). The Planck survey~\citep{2006astro.ph..4069T}, which is expected to report its first CMB results in early 2013,
will improve considerably
the statistical power of the CMB power spectrum measurements between $l=500$ where WMAP becomes noise limited, and $l=2000$ where Planck's poorer resolution
and sensitivity loses out to SPT despite the greatly extended sky coverage.
Beyond $l \sim 2000$, measurements start to be contaminated by
the unresolved background of extra-Galactic sources at all frequencies.

Given that the physics of the CMB is so well understood, the standard cosmological
model can be tested very precisely and its parameters determined to high
precision (see e.g.~\citealt{2011ApJS..192...18K,2011ApJ...743...28K}). Here, for brevity, we can highlight only three
examples.

\emph{Primordial power spectrum}: This affects the overall morphology of the
CMB power spectrum. Parameterising the spectrum of primordial
curvature perturbations as a power-law, $\mathcal{P}_{\mathcal{R}}(k) \propto
k^{n_s -1}$, WMAP constrains $n_s  = 0.963 \pm 0.014$ (assuming no contribution
to the CMB from gravitational waves;~\citealt{2011ApJS..192...16L}). This is beautifully consistent with
the inflationary prediction of a nearly scale-free spectrum ($n_s=1$). A departure
from scale-invariance is detected at almost the $3\sigma$ level and provides important
constraints on the dynamics of inflation [i.e.\ the slope and curvature of
$V(\phi)$]. The combination of the $n_s$ constraint and upper limits on the gravitational wave power spectrum (see Section~\ref{sec:gravitywaves}) already rule out several simple
inflation models.

\emph{Matter densities}: The relative heights of the acoustic peaks are
influenced by the physical densities of baryons and CDM. For example, increasing
the baryon fraction adds inertia but not pressure support to the plasma,
reducing the bulk modulus. This increases the overdensity at the midpoint of
the acoustic oscillations boosting the compressional peaks (1st, 3rd etc.).
Precise measurements of the physical baryon density, $\Omega_b h^2
= 0.02258\pm 0.00056$~\citep{2011ApJS..192...16L}, have been made via this route, nicely consistent with constraints
from big-bang nucleosynthesis.
This 3\% precision should improve to around 1\% with Planck data. Similarly,
the CDM density is measured to be $\Omega_c h^2 = 0.1109 \pm 0.0056$. This provides
inescapable evidence of the need for non-baryonic dark matter independently
of other lines of reasoning such as the clustering and internal kinematics of galaxies.

\emph{Curvature}: The angular scale of the acoustic peaks $r_s / d_A$ is now very
precisely measured. In standard models, the matter densities determined from the relative peak heights fully determine $r_s$, thus allowing an accurate measurement
of the angular-diameter distance to last scattering. This distance is very sensitive
to spatial curvature through its geometrical focusing effect, but this can always
be compensated by altering the radial distance to last-scattering (through
the Hubble constant $H_0$ or, equivalently, the dark energy density).
This leads to a \emph{geometric degeneracy}
whereby models with the same physical densities at high redshift, the same
primordial power spectrum, and the same angular-diameter distance to
last-scattering give almost identical angular power spectra. (An
example is given later in Fig.~\ref{fig:lensing_degeneracies}). The degeneracy
can be broken by adding other astrophysical distance measures such as
the Hubble constant, the angular-diameter distance at lower redshift (inferred
from the relic of the baryon acoustic oscillations -- BAO -- in the clustering of galaxies;~\citealt{2005ApJ...633..560E}) or the luminosity distance inferred from supernovae. An early,
important example of determining curvature via this route was from the 1998
flight of BOOMERanG~\citep{2000Natur.404..955D}. By precisely characterising the first acoustic peak,
the team were able to establish that space was flat at the 10\% level. More
recent measurements are consistent with flatness at the 0.5\% level~\citep{2011ApJS..192...18K},
strongly supporting one of the main predictions of inflationary cosmology.

\subsection{Polarization}

The other key CMB observable is polarization. Thomson scattering of unpolarized radiation with a quadrupole ($l=2$) anisotropy in its total intensity generates linear
polarization. The relevant epoch for the generation of polarization in the CMB
is around recombination since at early times scattering is too efficient to allow
a significant quadrupole to grow, while after recombination scatterings are very rare
(until the universe reionizes). The expected linear polarization from curvature perturbations has an r.m.s.\ of $5\, \mu\mathrm{K}$.

Linear polarization can be described by two Stokes parameters $Q$ and $U$. These
depend on a choice of basis and measure the difference in intensity transmitted
by linear polarizers aligned with the basis directions ($Q$) or at $45^\circ$ to them
($U$). While Stokes parameters provide a local, operational definition of polarization,
their coordinate dependence makes them rather inconvenient for cosmological
interpretation. Instead, linear polarization can be described in terms of two scalar
fields, $E$ and $B$~\citep{1997PhRvL..78.2054S,1997PhRvL..78.2058K}. The Stokes parameters are properly the components
in an orthonormal basis of a rank-2 symmetric, trace-free tensor which can be expressed in term of
second derivatives of $E$ and $B$ (neglecting sky curvature for simplicity, and
using Cartesian coordinates):
\begin{equation}
\left( \begin{array}{cc} Q & U \\
U & -Q 
\end{array} \right) \propto \left(\partial_i \partial_j - \frac{1}{2} \delta_{ij}
\nabla^2\right) E + \epsilon_{k(i} \partial_{j)} \partial_k B \, .
\end{equation}
This is analogous to decomposing a vector field into a gradient part ($E$)
and a divergence-free curl part ($B$). Note that $E$ and $B$ are non-local in
$Q$ and $U$.

The $E$-modes are scalars under parity but $B$-modes are pseudo-scalar. In
the absence of parity-violating physics, the two fields must be uncorrelated.
This leaves three non-zero polarization power spectra: the $E$- and $B$-mode
auto-correlations $C_l^E$ and $C_l^B$, and the cross-correlation $C_l^{TE}$ between
$E$ and the temperature anisotropies. The predicted angular power
spectra for inflationary curvature perturbations are shown in the left-hand
panel of Fig.~\ref{fig:scalar_tensor_spectra}. The main points to note are as follows:
(i) polarization is a small signal; (ii) $E$-mode polarization peaks on smaller scales
than the temperature, since it relies on diffusion in small-scale modes for its
generation; (iii) the acoustic peaks in $C_l^E$ are at the troughs of $C_l^T$
since the temperature quadrupole derives mostly from the plasma bulk velocity
which vanishes when the density is at an extremum; (iv) there is a `bump' in
the polarization on large scales generated by re-scattering once the universe
reionizes; and (v) by symmetry, curvature perturbations cannot generate $B$-mode polarization
except through second-order processes such as gravitational lensing (see
Section~\ref{sec:lensing}). This last point makes $B$-modes a potentially powerful
probe of gravitational waves; see Section~\ref{sec:gravitywaves}.

\begin{figure}
\begin{center}
\includegraphics[width=1.9in,angle=-90]{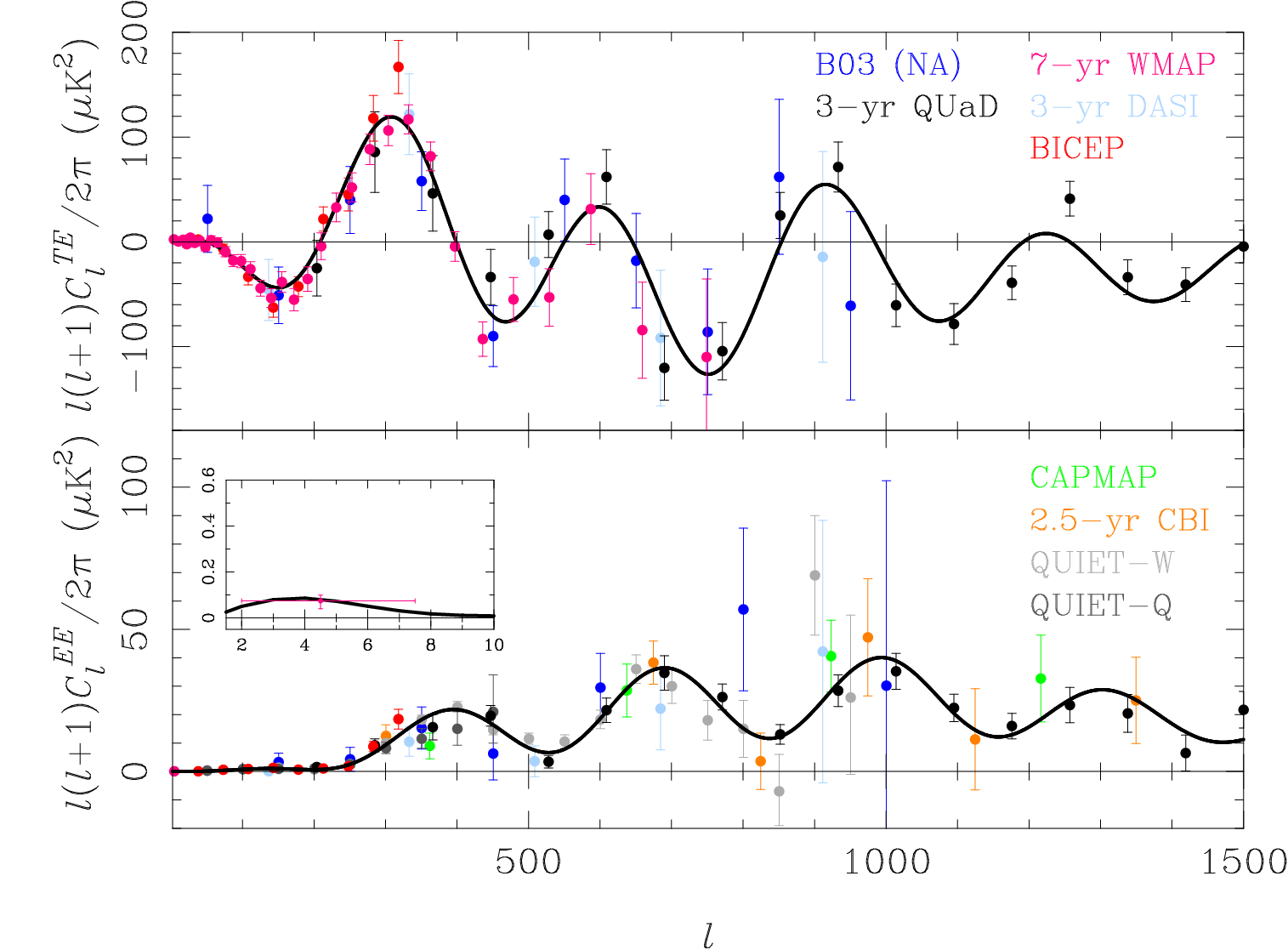}
\includegraphics[width=1.9in,angle=-90]{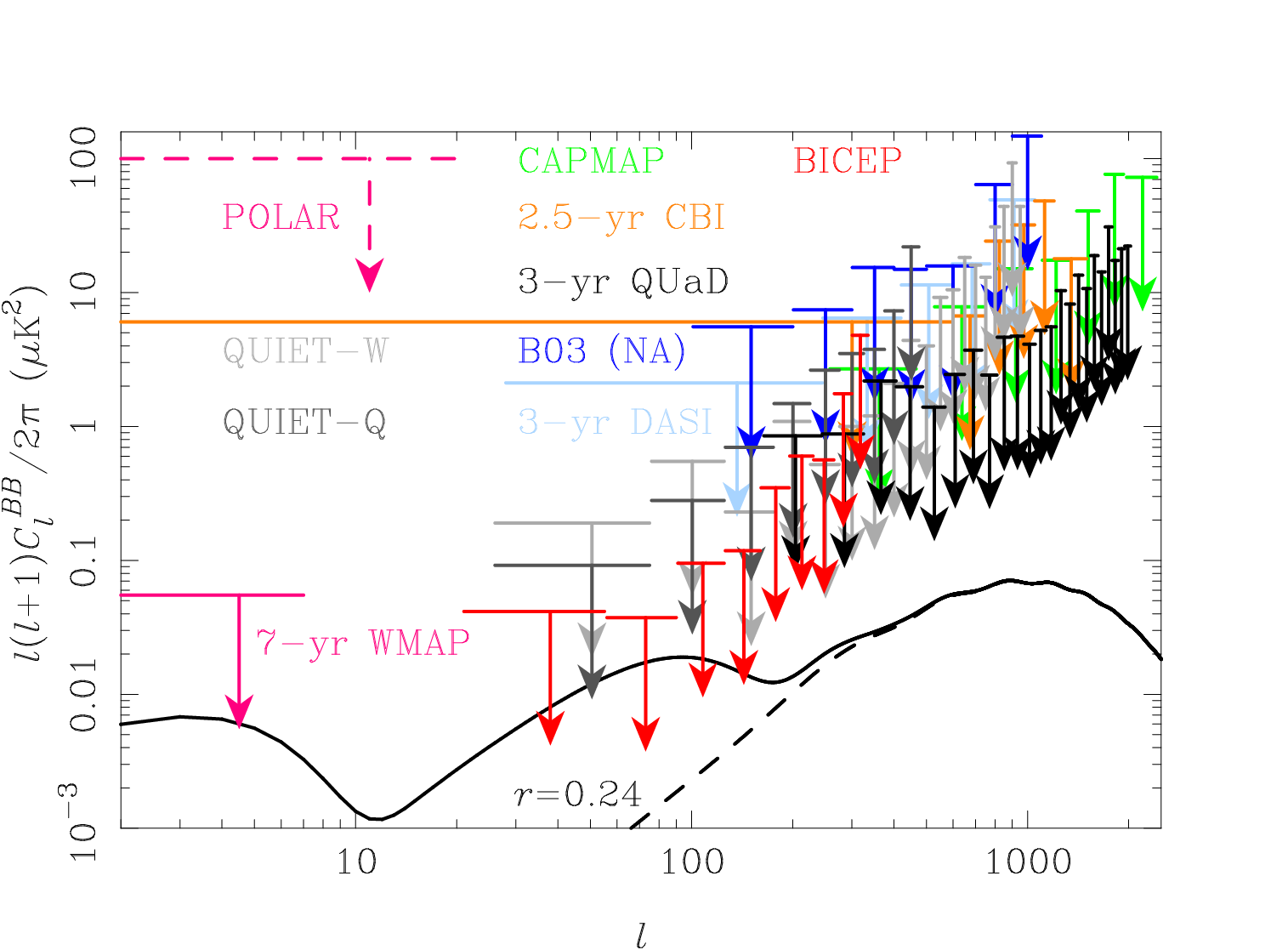}
\caption{\emph{Left:} current measurements of the polarization power spectra
$TE$ (top) and $EE$ (bottom) from WMAP7 (magenta;~\citealt{2011ApJS..192...16L}), QUaD  (black;~\citealt{2009ApJ...705..978B}),
BOOMERanG (blue;~\citealt{2006ApJ...647..833P,2006ApJ...647..813M}), DASI (cyan;~\citealt{2005ApJ...624...10L}), CAPMAP
(green;~\citealt{2008ApJ...684..771B}), CBI (orange;~\citealt{2007ApJ...660..976S}),
BICEP (red;~\citealt{2010ApJ...711.1123C}) and QUIET $W$-band (light grey;~\citealt{2012arXiv1207.5034Q})
and QUIET $Q$-band (dark grey;~\citealt{2011ApJ...741..111Q}). The lines are $\Lambda$CDM fits
to temperature and polarization data. \emph{Right:} current 95\% upper limits
on the $BB$ power spectrum including the constraint from POLAR (dashed
cyan;~\citealt{2001ApJ...560L...1K}). The dashed line is the contribution from gravitational waves
for $r=0.24$, the 95\% upper limit from fits to the temperature and $E$-mode
polarization data from WMAP7 combined with BAO and $H_0$ measurements~\citep{2011ApJS..192...18K},
and the solid line includes the contribution from gravitational lensing.}
\label{fig:polarization_measurements}
\end{center}
\end{figure}

Observations of CMB polarization are not yet as advanced as for the temperature
anisotropies. Current power spectrum measurements are shown in Fig.~\ref{fig:polarization_measurements}.  Antarctic experiments have played
a very significant role, including the first detection of CMB polarization by the DASI interferometer in 2002~\citep{2002Natur.420..772K}, and the current best characterisation of the
spectra by QUaD~\citep{2009ApJ...705..978B} and BICEP~\citep{2010ApJ...711.1123C}. The measurements are in
excellent agreement with expectations based on the temperature power spectrum,
providing an important consistency test. Moreover, through large-angle
$E$-modes, WMAP measures the optical depth to reionization to
be $\tau = 0.088\pm 0.015$~\citep{2011ApJS..192...16L}, providing an important integral constraint on astrophysical
models of reionization. Future $E$-mode polarization measurements will
tighten parameter constraints over those from the temperature anisotropies, particularly in non-standard models, and extend the angular range that can be reliably probed before foregrounds dominate. However, the real excitement over polarization is
the prospect of detecting the signature of gravitational waves
via $B$-modes and exploiting $B$-modes induced by weak lensing.

We end this section by emphasising that, despite the triumph of the standard
cosmological model in fitting essentially all cosmological data (with just six
parameters), the model raises several big questions. Did inflation happen?
What is the nature of dark matter? Why is the universe accelerating? In the following sections, we review how
ongoing and future CMB observations will help answer some of these questions.


\section{Gravitational waves and $B$-mode polarization}
\label{sec:gravitywaves}

Inflation naturally predicts the production of a stochastic background
of primordial gravitational waves accompanying the primordial density
perturbation~\citep{1979JETPL..30..682S}. The spectrum of gravitational
waves depends only on the expansion rate during inflation. Since this is
nearly constant during slow-roll inflation, with only a slow decrease,
the primordial spectrum $\mathcal{P}_h (k)$ should be well approximated
by a power-law with a slightly red spectrum. As the
Friedmann equation relates the expansion rate directly to the energy density
during inflation, a measurement of the gravitational wave power gives
directly the energy density and hence the \emph{energy scale} $E_{\mathrm{inf}}$
during inflation. It is conventional to express the amplitude of
$\mathcal{P}_h (k)$ in terms of its ratio to the power spectrum of
curvature perturbations $\mathcal{P}_{\mathcal{R}}(k)$ at a
cosmologically-relevant scale (often $k_0 = 0.002\,\mathrm{Mpc}^{-1}$). This
\emph{tensor-to-scalar} ratio $r$ is related to $E_{\mathrm{inf}}$ by
\begin{equation}
r = 8\times 10^{-3} (E_{\mathrm{inf}}/10^{16}\, \mathrm{GeV})^4 ,
\label{eq:r}
\end{equation}
where we have taken the scalar amplitude $\mathcal{P}_{\mathcal{R}}(k_0)= 2.36 \times 10^{-9}$. Note that
$r \sim 10^{-2}$ for inflation occurring around the GUT scale,
$E_{\mathrm{inf}} \sim 10^{16}\, \mathrm{GeV}$.


Gravitational waves damp away due to the expansion of the universe when their
wavelength is smaller than the Hubble radius. The best prospect for
detection is therefore via the CMB which is sensitive to early times (after
last-scattering) and large scales. Gravitational waves generate CMB temperature
anisotropies due to the integrated effect of the anisotropic expansion they
induce along the line of sight; see Fig.~\ref{fig:scalar_tensor_spectra}. However, the signal is
limited to large angular scales, $l < 60$,
corresponding to gravitational waves with wavelengths larger than the Hubble
radius at last-scattering. On such scales, chance upwards fluctuations in the
temperature anisotropies from curvature perturbations due to cosmic variance
limit our ability to measure $r$. In the optimistic scenario that all other
cosmological parameters are know, cosmic variance gives a $1\sigma$ error
on $r$ of $0.07$ from the temperature anisotropies alone. In practice,
degeneracies make the CMB-only limit a little worse:
e.g.\ $r < 0.21$ (at 95\% confidence)
from WMAP7+SPT~\citep{2011ApJ...743...28K}, improving on
$r<0.36$ from WMAP7 alone~\citep{2011ApJS..192...18K}. 
Gravitational waves also leave an imprint in the linear polarization of
the CMB. Significantly, they generate $E$- \emph{and} $B$-modes
with roughly equal power, unlike curvature perturbations which only
generate $B$-modes at second order through gravitational lensing. 
In principle, $B$-mode measurements of $r$ can do much better than inferences
from the temperature or $E$-mode polarization since the former is only limited
by the cosmic variance of the lens-induced $B$-modes\footnote{In principle,
the cosmic variance from lensing can even be removed by ``delensing'' the
observed $Q$ and $U$ maps with a reconstruction of the lensing deflection
field~\citep{2004PhRvD..69d3005S}. The latter can be obtained from the CMB itself with high-resolution polarization observations; see Section~\ref{sec:lensing}.}. The problem
is that the $B$-mode signal is very small (see Fig.~\ref{fig:scalar_tensor_spectra}); the limit
$r < 0.24$ implies that the r.m.s.\ from gravitational waves
is less than $200\,\mathrm{nK}$. The measurement therefore requires exquisite
sensitivity and control of systematic effects to maintain polarization purity,
and careful rejection of polarized emission from our Galaxy.

Current upper limits on the $B$-mode power spectrum are shown in
Fig.~\ref{fig:polarization_measurements}. The best constraints over nearly the full range of scales
come from BICEP~\citep{2010ApJ...711.1123C} at degree
scales and QUaD~\citep{2009ApJ...705..978B} on smaller scales. The BICEP constraint of
$r < 0.73$ (95\% confidence) is not yet competitive with that from
the temperature anisotropies, although it is rather less model dependent.
There are two main scales to attempt detection of $B$-modes from gravitational
waves: $l< 10$ where the signal is generated by scattering at reionization,
and $l \sim 100$ where the signal from scattering around recombination peaks.
The reionization signal needs a nearly full-sky survey and so broad
frequency coverage to remove Galactic emission which is dominant over
most of the sky. The best near-term constraints on these scales will come from
Planck with forecasts indicating $r < 0.05$ may be achievable~\citep{2009JCAP...06..011E}.
The signal
from recombination can be constrained by targeting clean, connected
regions of the sky (typically around $1000\,\mathrm{deg}^2$)
in areas of low Galactic emission. By good fortune,
one of the cleanest such regions is accessible from Antarctica. A series of
results from BICEP's successors, BICEP2, Keck and POLAR (see contributions
from Pryke and Kuo in this volume), as well as the balloon-borne SPIDER~\citep{2010SPIE.7741E..46F} and several
experiments based in Atacama, are eagerly
anticipated over the next five years. These should push down the errors on
$r$ to around $0.01$. This is an interesting target for inflationary physics
since the signal from a large class of simple models -- ``large-field'' such
as monomial potentials -- would be detectable. 
Looking further ahead,
the constraint on $r$ could plausibly be improved to the $10^{-4}$--$10^{-3}$
level with a future polarization satellite.

\section{Weak gravitational lensing of the CMB}
\label{sec:lensing}

The fluctuations in the temperature of the CMB are mostly imprinted at the epoch of last scattering. However, CMB photons undergo small gravitational deflections
due to the clumpy distribution of matter (weak gravitational lensing) as they
propagate from last-scattering to the present epoch. The r.m.s.\ deflection is only $2.7\,\mathrm{arcmin}$ but is coherent over several degrees. The lensing effect is similar to seeing the CMB fluctuations from the last-scattering surface through patterned glass, and subtly distorts their statistics. With telescope
resolution of a few arcminutes or better, these distortions can be detected and
used to reconstruct the lensing deflection. This opens up a new cosmological probe of structure formation at epochs and scales that are difficult to access
with more direct probes (such as galaxy clustering). Lensing is an emerging
field in observational CMB research and results from the SPT are at the
forefront of this.

\begin{figure}
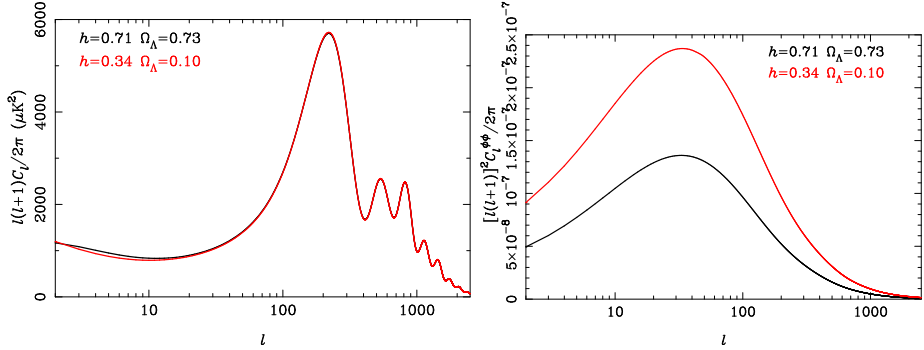

\begin{center}
\includegraphics[width=1.8in,angle=-90]{Cls_nonlens_degen.ps}
\includegraphics[width=1.8in,angle=-90]{cldd_degen.ps}
\caption{Breaking the angular-diameter distance degeneracy with CMB lensing.
The unlensed (and unobservable!) temperature power spectra (left) for the
standard $\Lambda$CDM model (black) and a closed model with low $H_0$ to
match the angular scale of the acoustic peaks (red) are very nearly degenerate.
The degeneracy is broken
in the power spectrum of the lensing deflection angle (right) since matter
is more clustered at late times in the model with low $H_0$.}
\label{fig:lensing_degeneracies}
\end{center}
\end{figure}

Weak lensing has several important effects on the CMB;
see~\citet{2006PhR...429....1L} for a detailed review.
Magnification and demagnification of the acoustic-scale features leads to
a smoothing of the acoustic peaks, reaching the 10\% level at $l > 2000$
in temperature, and rather larger in $E$-mode polarization.
On smaller scales, for which the unlensed CMB is very smooth, lensing
generates small-scale power that dominates the
primary anisotropies for $l > 4000$. The lens remapping moves around the
polarization amplitude while preserving the direction, generating $B$-modes
from the primary $E$-modes with an almost white spectrum for $l\ll 1000$;
see Fig.~\ref{fig:scalar_tensor_spectra}. As noted in Section~\ref{sec:gravitywaves}, this will become
an important source of confusion for CMB searches for gravitational waves.
Finally, lensing introduces four-point non-Gaussianity with a very specific and predictable shape from which the full angular power spectrum $l(l+1)C_l^{\phi\phi}$ of the lensing deflections can be reconstructed\footnote{In linear theory, the deflection is the gradient of the \emph{lensing potential} $\phi$. The lensing potential is an integrated measure of the gravitational potential along the line of sight.}. Through these lensing effects, the CMB is sensitive to parameters that have degenerate effects in the primary anisotropies. For example, Fig.~\ref{fig:lensing_degeneracies} compares the unlensed temperature power spectra and the deflection power spectra for the standard $\Lambda$CDM model and a closed model with low
$H_0$. These models lie along the geometric degeneracy of the unlensed CMB power spectra. However, the deflection spectra are quite different since matter is more clustered at late times in the
low-$H_0$ model. Other parameters that benefit similarly from lensing information include sub-eV neutrino masses and early dark energy.

\begin{figure}
\begin{center}
\includegraphics[width=2.8in,angle=0]{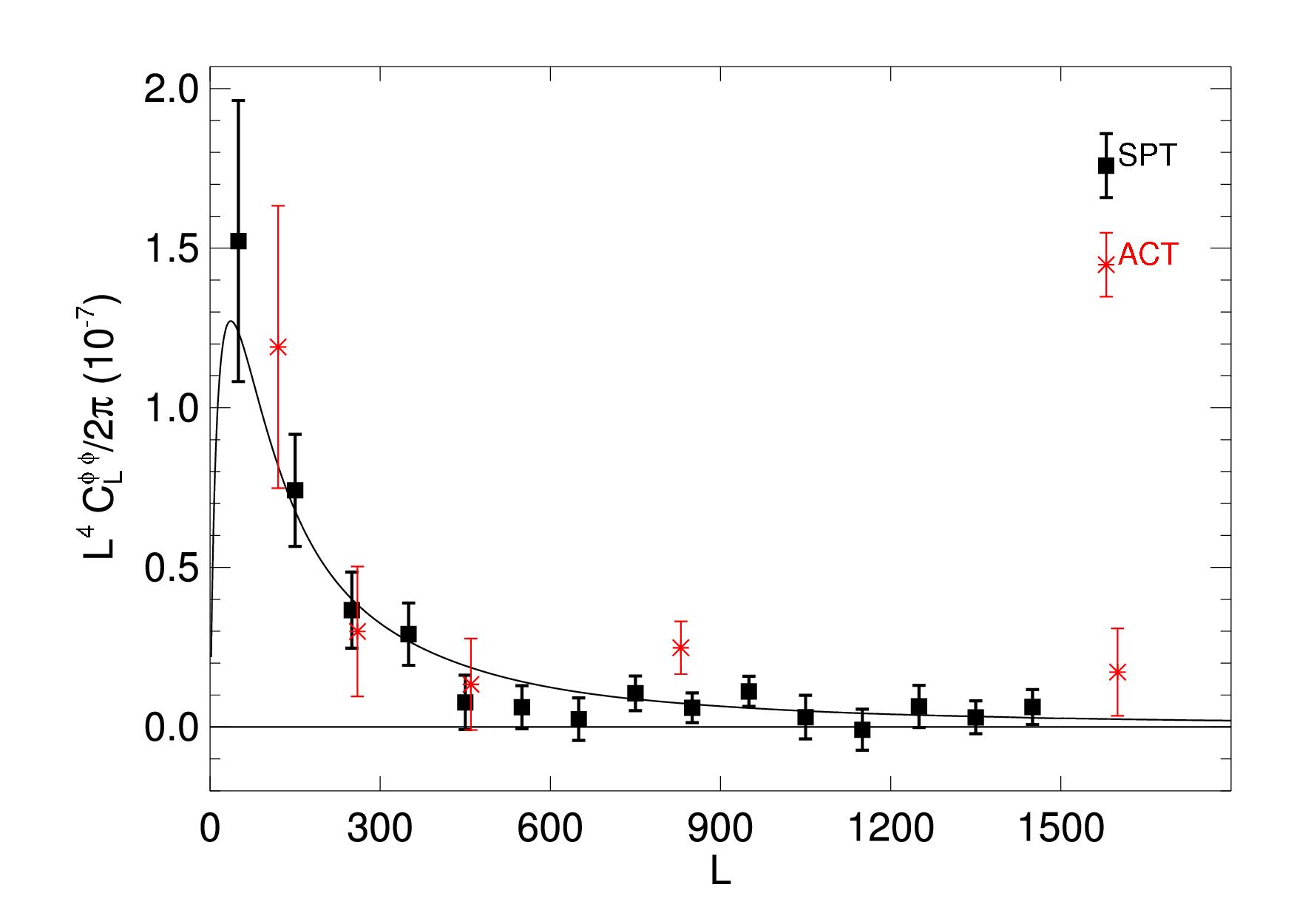}
\includegraphics[width=2.4in,angle=0]{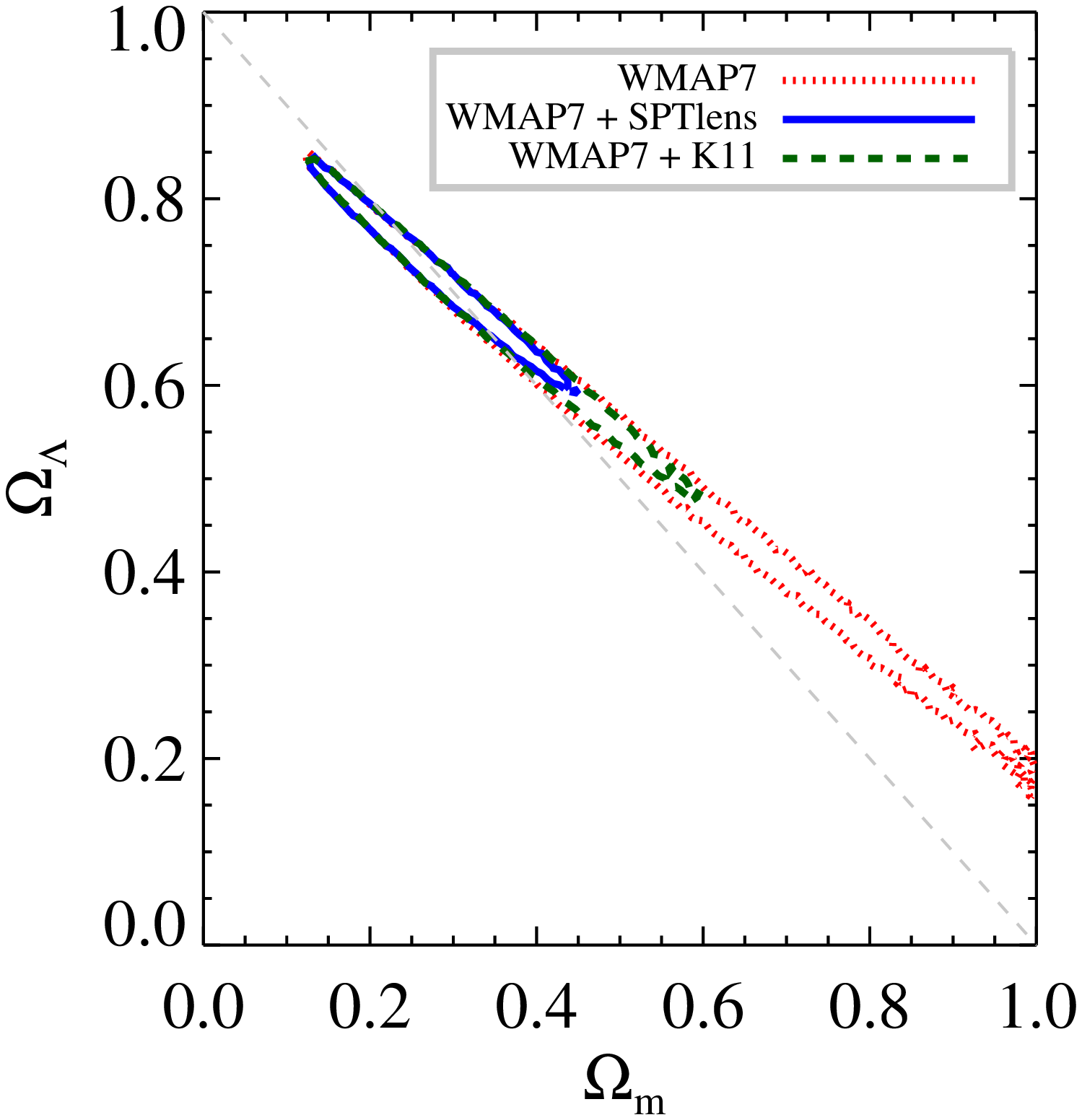}
\caption{\emph{Left:} Current measurements of the lensing deflection
power spectrum from SPT (black;~\citealt{2012ApJ...756..142V}) and ACT (red;~\citealt{2011PhRvL.107b1301D}).
The solid line is a $\Lambda$CDM fit to CMB temperature and polarization data,
but not to the lensing data. \emph{Right:} 95\% confidence regions in the
$\Omega_\Lambda$--$\Omega_m$ plane for $\Lambda$CDM models with curvature.
The geometric degeneracy is evident in the WMAP7-alone
constraints (red), but the tail of low-$H_0$ closed models is cut-off
by the higher-resolution SPT data which is sensitive to the lensing effect
in the temperature power spectrum (green;~\citealt{2011ApJ...743...28K}).
Even tighter constraints are
obtained by combining the SPT lens reconstruction from the left-hand figure
with the WMAP7 data (blue). Figure reproduced with permission
from~\cite{2012ApJ...756..142V}.}
\label{fig:lensing_measurements}
\end{center}
\end{figure}

The first measurements of the deflection power spectrum from the
four-point function of the temperature anisotropies have recently been
reported by the Atacama Cosmology Telescope (ACT;~\citealt{2011PhRvL.107b1301D}) and SPT~\citep{2012ApJ...756..142V}. These are in excellent agreement
with expectations for the standard $\Lambda$CDM model; see Fig.~\ref{fig:lensing_measurements}.
The current SPT measurements constitute a $6.3\sigma$ detection but they are
from only $590\,\mathrm{deg}^2$ of sky. The significance can be expected to
increase several-fold with the analysis of the full $2500\,\mathrm{deg}^2$
survey, similar
to what should be achieved with Planck. Already, combining the SPT lens
reconstruction with the temperature power spectrum from WMAP breaks
the geometric degeneracy in $\Lambda$CDM models with curvature giving
a significant detection of dark energy from the CMB alone; see the right-hand panel of Fig.~\ref{fig:lensing_measurements}.
The degeneracy is also broken by the effect of lensing on the high-$l$
temperature power spectrum itself, as measured, for example, by SPT~\citep{2011ApJ...743...28K}.

Lens reconstruction from the CMB temperature suffers from statistical noise due to chance correlations in the unlensed CMB that mimic the effect of lensing. This is such that temperature reconstructions will never give cosmic-variance-limited measurements of the deflection power spectrum for multipoles $l > 100$. Polarization measurements are very helpful here~\citep{2002ApJ...574..566H}, since they intrinsically have more small-scale power and the $B$-mode of polarization is not confused by primary anisotropies. In principle, polarization can provide cosmic-variance limited reconstructions to multipoles $l \approx 500$, i.e.\ on all scales where linear theory applies. For this reason, lens reconstruction from polarization is an important science goal for the polarization upgrades to the SPT~\citep{2009AIPC.1185..511M} and ACT~\citep{2010SPIE.7741E..51N}, as well as proposed successors to the Planck satellite~\citep{2008arXiv0805.4207B,2009arXiv0906.1188B,2011arXiv1102.2181T}.

\section{Outlook}

The future of CMB observations lies on several fronts.
Precise polarization measurements on large scales will greatly improve limits
on the stochastic background of gravitational waves predicted from inflation
in the early universe. Wide-area, high-resolution temperature and polarization
measurements will allow precise reconstruction of the gravitational-lensing effect
in the CMB and provide a new window to the large-scale clustering of matter
around redshift two.
In addition, arcminute-scale observations will provide catalogues of thousands
of galaxy clusters over a broad redshift range and with well-understood selection functions, and measure the Doppler signatures from the bulk flows of matter
in the post-reionization universe.  The cluster catalogues will be used to probe the growth of structure and evolution
of the volume element to high redshift. These programmes address directly many of the
outstanding issues raised by the standard cosmological model, such as
the physics of inflation and the cause of the current accelerated expansion.


\end{document}